\documentclass[aps,prd,twocolumn,
amssymb,amsmath,showpacs,a4paper,superscriptaddress,nofootinbib,longbibliography]{revtex4-2}
\usepackage{graphicx}
\usepackage{braket}
\usepackage{amsfonts}
\usepackage{amsmath}
\usepackage{bbold}
\usepackage{hyperref}
\usepackage{adjustbox}
\usepackage{float}
\usepackage[nameinlink]{cleveref}
\Crefname{figure}{Fig.}{Figs.}
\usepackage[table,xcdraw]{xcolor}
\newcommand{\beq}{\begin{equation}}
	\newcommand{\eeq}{\end{equation}}
\newcommand{\bea}{\begin{eqnarray}}
	\newcommand{\eea}{\end{eqnarray}}
\newcommand{\bit}{\begin{itemize}}
	\newcommand{\eit}{\end{itemize}}
\newcommand{\ben}{\begin{enumerate}}
	\newcommand{\een}{\end{enumerate}}
\newcommand{\nn}{\nonumber}

\begin{document}
	
    \title{Probing the Unstable Spectrum of Schwarzschild-like Black Holes}
	
	\author{Pedro Henrique Croti Siqueira}
	\email{pedro.croti@ufabc.edu.br}
       \affiliation{Centro de Matem\'atica, Computa\c c\~ao e Cogni\c c\~ao, Universidade Federal do ABC (UFABC), 09210-170 Santo Andr\'e, S\~ao Paulo, Brazil}
	\affiliation{Centro de Ci\^encias Naturais e Humanas, Universidade Federal do ABC (UFABC), 09210-170 Santo Andr\'e, S\~ao Paulo, Brazil}
	\author{Lucas Tobias de Paula}
	\email{tobias.l@ufabc.edu.br}
        \affiliation{Centro de Matem\'atica, Computa\c c\~ao e Cogni\c c\~ao, Universidade Federal do ABC (UFABC), 09210-170 Santo Andr\'e, S\~ao Paulo, Brazil}
	\affiliation{Centro de Ci\^encias Naturais e Humanas, Universidade Federal do ABC (UFABC), 09210-170 Santo Andr\'e, S\~ao Paulo, Brazil}
	\author{Rodrigo Panosso Macedo}
	\email{rodrigo.macedo@nbi.ku.dk}
	\affiliation{Niels Bohr International Academy, Niels Bohr Institute, Blegdamsvej 17, 2100 Copenhagen, Denmark}
	\author{Maur\'icio Richartz}
	\email{mauricio.richartz@ufabc.edu.br}
	\affiliation{Centro de Matem\'atica, Computa\c c\~ao e Cogni\c c\~ao, Universidade Federal do ABC (UFABC), 09210-170 Santo Andr\'e, S\~ao Paulo, Brazil}

\begin{abstract}
 We investigate the pseudospectrum of a Schwarzschild-like spacetime within the framework of black hole perturbation theory to analyze a counterintuitive assertion regarding the instability of quasinormal modes. Recent findings suggest that random perturbations to the effective potential associated with gravitational waves may enhance the stability of the underlying wave operator, thereby yielding a stable spectrum of randomly displaced quasinormal modes. Given the unphysical nature of such random perturbations, this work examines these findings within a spacetime that inherently exhibits a perturbed quasinormal spectrum. We find that, in contrast to the QNM spectrum of the Schwarzschild spacetime under random perturbations, the quasinormal spectrum of a Schwarzschild-like black hole deformed through a physically motivated implementation of the Rezzolla-Zhidenko parametrization is unstable. In particular, we show that the pseudospectra of these Schwarzschild-like black holes do not display the typical features associated with wave operators that yield stable quasinormal spectra.  We corroborate our findings by computing the quasinormal spectra when additional ({\em ad-hoc}) deformations are added to the effective potential of the Rezzolla-Zhidenko black hole. We also argue that when multiple perturbation sources are present, identifying the origin of the instability may be difficult.
    \end{abstract}
	\maketitle
	%%%%%%%%%%%%%%%%%%%%%%%%%%%%%%%%%%%%%%%%%%%%%%%%%%%%%%%%%%%%%%%%%%%%%%%%%%%%%%%%%%%%%%%%%%%%%%%%
	
	\section{Introduction}	
In recent decades, the study of black holes and their spectral properties has become a central area of research in theoretical physics, largely driven
by the pivotal gravitational wave observations conducted by LIGO, Virgo, and KAGRA~\cite{LIGO2016,KAGRA:2021vkt,LIGOScientific:2024elc}. These experiments have opened a new avenue to study black holes through spectroscopy \cite{Dreyer:2003bv,Berti:2005ys,Berti:2016lat}, enabling the use of gravitational wave signatures, from systems such as black hole mergers, to infer details about the geometry and physics of the underlying spacetime. A key aspect of black hole spectroscopy is the fact that gravitational waves emitted by a perturbed black hole are accurately described by a superposition of exponentially damped sinusoids whose frequencies correspond to the so-called quasinormal modes (QNMs)~\cite{Kokkotas:1999bd,Berti;2009,Zhidenko;2011}.

The black hole spectroscopy program is fundamentally based on linearized calculations of the vacuum black hole spectrum within general relativity~\cite{Kokkotas:1999bd,Berti;2009,Zhidenko;2011,Barausse:2014tra,Dreyer:2003bv,Berti:2005ys}. This framework inherently assumes that the spectrum exhibits a reasonable degree of physical robustness against perturbations in the system. In simple and intuitive terms, it is generally expected that small modifications to the black hole geometry will not result in significant changes to its QNM spectrum. Due to the loss of energy in form of gravitational waves through the event horizon and out to the wave zone, however, black hole perturbation theory lies within the realm of non-Hermitian physics \cite{Ashida:2020dkc,PhysRevX.11.031003, Dias:2022oqm,Motohashi:2024fwt,Cavalcante:2024swt}. 
Therefore, key mathematical theorems ensuring the stability of Hermitian systems are not applicable to the QNM spectrum, and black holes may, in fact, exhibit an unstable spectrum.
This instability phenomenon, initially observed in early studies of the field~\cite{Vishveshwara:1996jgz,Aguirregabiria:1996zy,Nollert:1996rf,Nollert:1998ys}, has recently been incorporated into a more formal framework~\cite{PhysRevX.11.031003,Jaramillo:2021tmt} that imports tools from the theory of non-self-adjoint operators~\cite{trefethen2005spectra, Sjostrand2019} into gravitational research~\cite{PhysRevX.11.031003}. Specifically, the concept of the pseudospectrum~\cite{trefethen2005spectra} offers a robust tool to diagnose, characterize and visualize the QNM spectral instability. Moreover, it enables a non-modal analysis~\cite{Jaramillo:2022kuv}, providing deeper insights into the stability properties of the system. 

Broadly speaking, the pseudospectrum can be understood as a topographic map on the complex plane, 
whose peaks coincide with the complex eigenvalues of the unperturbed operator, i.e., the QNMs. Spectral properties are effectively visualized through the contour lines of this topographic map, which correspond to the level sets associated with a constant parameter $\varepsilon$.
These level sets, defining the  $\varepsilon$-pseudospectra, delineate regions in the complex plane where eigenvalues may “migrate” under perturbations of magnitude $\varepsilon$ (quantitative assessments must be performed carefully~\cite{PhysRevX.11.031003,Boyanov:2023qqf,Cownden:2023dam, Boyanov:2024fgc,Besson:2024adi}).
The spectrum of the unperturbed system corresponds to the limit  $\varepsilon \to 0$. Consequently, concentric contour lines with steep gradients around eigenvalues indicate spectral stability, while widely spaced contour lines with shallow gradients extending far from eigenvalues are indicative of spectral instability.
   
A pseudospectral analysis has been applied effectively to several spacetimes and propagating fields~\cite{PhysRevX.11.031003,Destounis:2021lum,Jaramillo:2022kuv,Cao:2024oud,Sarkar:2023rhp,Destounis:2023nmb,Luo:2024dxl,Arean:2023ejh,Cownden:2023dam,Boyanov:2023qqf,Chen:2024mon,Cai:2025irl,dePaula:2025fqt}, with all scenarios diagnosing spectral instabilities. The introduction of small modifications to the operator associated with the QNM problem allows one to verify the instability {\em a posteriori}. A crucial question lies in determining which perturbations trigger such instabilities \cite{PhysRevX.11.031003} and whether they are physically relevant or not~\cite{Jaramillo:2022kuv,Cardoso:2024mrw}. 
Nevertheless, even unphysical perturbations provide valuable insights into the properties of the operator and the resulting QNM spectra from a proof-of-principle perspective. For instance, random perturbations applied to the black hole potential in the underlying wave equation were initially employed in agnostic studies of QNM spectral instabilities. Despite their unphysical nature, such perturbations are known to maximize the displacement of QNMs along the $ \varepsilon$-pseudospectra~\cite{PhysRevX.11.031003}.

Interestingly, Ref.~\cite{PhysRevX.11.031003} also reports a counterintuitive result: the addition of a random perturbation to a spectrally unstable operator 
results in a new, randomly displaced QNM spectrum, which is itself stable to further perturbations. In other words, the pseudospectrum associated with the randomly perturbed operator exhibits the typical features of a stable spectra (concentric circles around the new eigenvalues, followed by flat patterns). As discussed in \cite{PhysRevX.11.031003}, this results from an improvement in the analytical behavior of the operator's resolvent (see definition in Sec.~\ref{sec:pseudospectrum}). 

%%%
In this work, we investigate the robustness of the above assertion in scenarios where the QNM instability is triggered by a more realistic spacetime deformation, rather than an {\em ad-hoc} random perturbation added to the wave equation potential. Specifically, we seek to answer the following question: How stable is a QNM spectrum already destabilized by a physically motivated deviation from Schwarzschild? To address this question, we employ the pseudospectrum approach to analyze the stability of the QNM spectrum associated with the Rezzolla-Zhidenko (RZ) metric~\cite{Rezzolla2014}. Alongside the analogous parametrization designed for axisymmetric spacetimes~\cite{Konoplya:2016jvv}, the RZ metric serves as a valuable tool for probing black holes that exhibit slight deviations from vacuum general relativity predictions~\cite{Volkel:2019muj,Konoplya:2020hyk,Konoplya:2022tvv,konoplya2022overtonesprobeeventhorizon,Stefano2021,Siqueira:2022tbc}. Moreover, it has been observed that such slight deviations can also trigger QNM spectral instabilities~\cite{Konoplya:2020hyk,Konoplya:2022tvv,konoplya2022overtonesprobeeventhorizon,Cardoso:2024mrw}, with overtones associated with the RZ metric differing significantly from their Schwarzschild counterparts. Given the broad and general character of the RZ parametrization, we restrict ourselves to the physically motivated configuration introduced in Ref.~\cite{Cardoso:2024mrw}.

Our paper is organized as follows. In Sec.~II we define the Schwarzschild-like spacetime described by the RZ metric that we consider in this work. In Sec.~III, after defining the hyperboloidal coordinates, we introduce the notion of the pseudoespectrum and the numerical methods employed to calculate it. Our main results are presented in Sec.~IV, where we investigate the QNM spectrum and the pseudospectrum of the RZ spacetime. We conclude with our final remarks in Sec.~V. Throughout this work we use units in which $G = c = \hbar = 1$.

%%%%%%%%%%%%%%%%%%%%%%%%%%%%%%%%%%%%%%%%%%%%%%%%%%%%%%%%%%%%%%%%%%%%%%%%%%%%%%%%%%%%%%%%%%%%%%%%
	%
	\section{The Schwarzschild-like spacetime} 
	\label{Sec:def_krz}
	\subsection{Spacetime metric}
	The most general line element describing a static, spherically symmetric spacetime reads
	\begin{equation}
		ds^2 = -a(r) dt^{2} + \dfrac{1}{b(r)} dr^{2} + r^{2} (d \theta^{2} + \sin^{2} \theta d \phi^{2}),
		\label{eq:krz-metric}
	\end{equation}
where $a$ and $b$ are arbitrary functions of the Schwarzschild radial coordinate $r$. By calculating the Einstein tensor from the line element above, one identifies an anisotropic matter content characterized by the stress tensor $ T^\mu{}_\nu =\mathrm{diag}\left(-\rho, p_{r}, p_{t}, p_{t} \right)$, where the mass density $\rho$, the radial pressure $p_r$, and the tangential pressure $p_t$ are 
    \bea
	 && \rho = \dfrac{m'(r)}{4 \pi r^2}, \\
  &&	p_{r} \! = \dfrac{b(r) a'(r)}{8 \pi r a(r)} - \dfrac{2 m(r)}{8 \pi r^3}, \\
  && p_t \! = \dfrac{b(r)}{32 \pi} \bigg\{ \frac{2 \left[  a(r) b(r) \right]'}{r a(r) b(r)}  
	   +  \frac{\left[ a'(r) b(r) \right]'}{a(r) b(r)} \! + \! \left[\dfrac{a'(r)}{a(r)} \right]' \!  \bigg\}.
	\eea
	The mass function $m(r)$ appearing in the above expressions is given by
	\begin{equation}
		m(r) = \frac{r\left[1-b(r)\right]}{2}.
	\end{equation}

	In this work, we fix the metric functions $a(r)$ and $b(r)$ in terms of the RZ parametrization introduced in \cite{Rezzolla2014}. The RZ scheme provides a generic representation of spherically symmetric, static and asymptotically flat spacetimes by defining
	\begin{align}
		a(r) = x A(x), \quad b(r) = \dfrac{x A(x)}{B(x)^{2}},		\label{eq:2}
	\end{align}
    where $x$ is a dimensionless variable that compactifies the radial coordinate. For instance, if the spacetime metric describes a black hole and $r_{0}$ denotes the position of its event horizon, the transformation between the standard Schwarzschild coordinate $r$ and the compactified coordinate $x$ is
	\begin{equation} \label{radicompact}
		x = \left( 1 - \dfrac{r_{0}}{r}\right).
	\end{equation}

The RZ scheme expresses the functions $A(x)$ and $B(x)$ in terms of a countably infinite set of dimensionless parameters (denoted by $\epsilon$, $a_{i}$ and $b_{i}$, with $i \in \mathbb{N}$)\footnote{In this work, $\epsilon$ denotes a parameter in the RZ framework, while $\varepsilon$ is used to define the pseudospectrum.  The notation and context ensure unambiguous distinction between the two.}

	\begin{align}
		A(x) &= 1 - \epsilon \left(1 - x\right) + \left( a_{0} - \epsilon \right)(1-x)^{2} + \tilde{A}(x)(1-x)^{3}, \nn \\
		B(x) &= 1 + b_{0}(1-x) + \tilde{B}(x)(1-x)^{2}, \label{eq:Bdex}
	\end{align}
    where
    \begin{equation}
		\! \! \tilde{A}(x) = \dfrac{a_{1}}{1+\dfrac{a_{2} x}{1+\dfrac{a_{3}}{1+ \cdots}}}, \ \ \ \ \tilde{B}(x) = \dfrac{b_{1}}{1+\dfrac{b_{2} x}{1+\dfrac{b_{3}}{1+ \cdots}}}. \label{eq:AeBtilde}
	\end{equation}
The Schwarzschild spacetime is recovered when $\epsilon = a_{0} = a_{1} = b_{0} = b_{1} = 0$. Hence, the RZ parameters that define \eqref{eq:krz-metric} 
quantify deviations from the Schwarzschild metric.

	Note that relevant physical quantities can be computed using the RZ parametrization. For instance, the ADM mass and the Hawking temperature are given, respectively, by
	\beq
	M_{\rm ADM} = \dfrac{r_0(1 + 2 b_0 + \epsilon)}{2},
	\eeq
    and
	\begin{equation}
		\label{temp}
		T_{H} = \dfrac{(1+a_{0}+a_{1}-2 \epsilon)}{4 \pi r_{0} \left(1 + b_{0} + b_{1}\right)}.
	\end{equation}
	At the event horizon, the energy density, along with the radial and tangential pressures, is expressed as
       \bea
	\label{stresstensor}
	 && \left.  8 \pi r_{0}^{2} \rho\right|_{r_0} = \left.- 8 \pi r_{0}^{2} p_{r}\right|_{r_0} =  1- \dfrac{\left(1 + a_{0} + a_{1} - 2\epsilon\right)}{\left(1 + b_{0} + b_{1}\right)^{2}} ,   \\
	 && \left. 16 \pi r_0^2 p_t\right|_{r_0} = \frac{1}{ (b_0+b_1+1)^3} \bigg\{ a_0 [-3 b_0+b_1 (b_2-2)-4] \nn \\
	&&  \ \ \ \ \ \ \ + a_1 [-2 a_2 (b_0+b_1+1)-5 b_0+b_1
		(b_2-4)-6] \nn \\ 
	&& \ \ \ \ \ \ \ \ + 2 \epsilon  (2 b_0-b_1 b_2+b_1+3)+ b_0+b_1
		(b_2+2) \bigg\}.  \label{stresstensor2}
	\eea
	
	  Gravitational perturbations of the RZ spacetime are analyzed under the assumption that the fluid permeating the spacetime is non-dissipative. Thus, they can be expanded in harmonics with index $ \ell \in \mathbb{Z} $ and decomposed into axial and polar sectors. We will focus on the axial sector, for which the perturbations of frequency $\omega$ are governed by \cite{Cardoso:2021wlq,Cardoso:2024mrw}
	\begin{equation}
    \label{eq:QNM_operator_SchwarCoord}
		\dfrac{d^2\psi(r)}{dr^2_{*}} + \bigg[\omega^2 - V(r)\bigg]\psi(r) = 0,
	\end{equation} 
	where $r_{*}$ is the tortoise coordinate, defined by
	\begin{equation}
        \label{eq:def_r_*}
		\dfrac{dr_{*}}{dr} = \dfrac{1}{f(r)}, \quad f(r) = \sqrt{a(r) b(r)},
	\end{equation} 
	and the effective potential $V(r)$ is given by
	\begin{equation}
		V(r) = a(r) \left[\dfrac{l(l+1)}{r^{2}}  - \dfrac{6m(r)}{r^{3}} + \dfrac{m'(r)}{r^{2}}\right].
	\end{equation}
    For a Schwarzschild black hole, this is the so-called Regge-Wheeler potential. The boundary conditions for \eqref{eq:QNM_operator_SchwarCoord}, associated with the QNM spectrum, are
    \beq
    \label{eq:BC}
    \psi(r)\sim e^{\pm i \omega r_*}, \quad r_* \rightarrow \pm \infty,
    \eeq
    ensuring that energy flows into the black hole horizon and into the infinitely far wave zone.

   \begin{figure*}[!t]
		\centering
		\includegraphics[width = 1. \columnwidth]{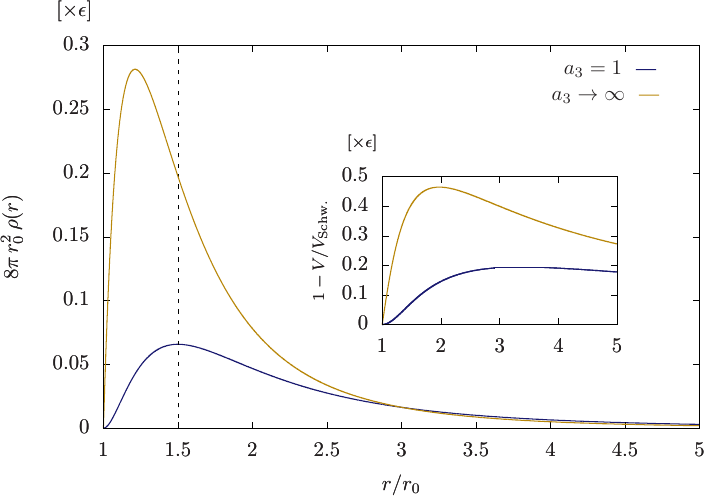}
        \includegraphics[width = 0.975 \columnwidth]{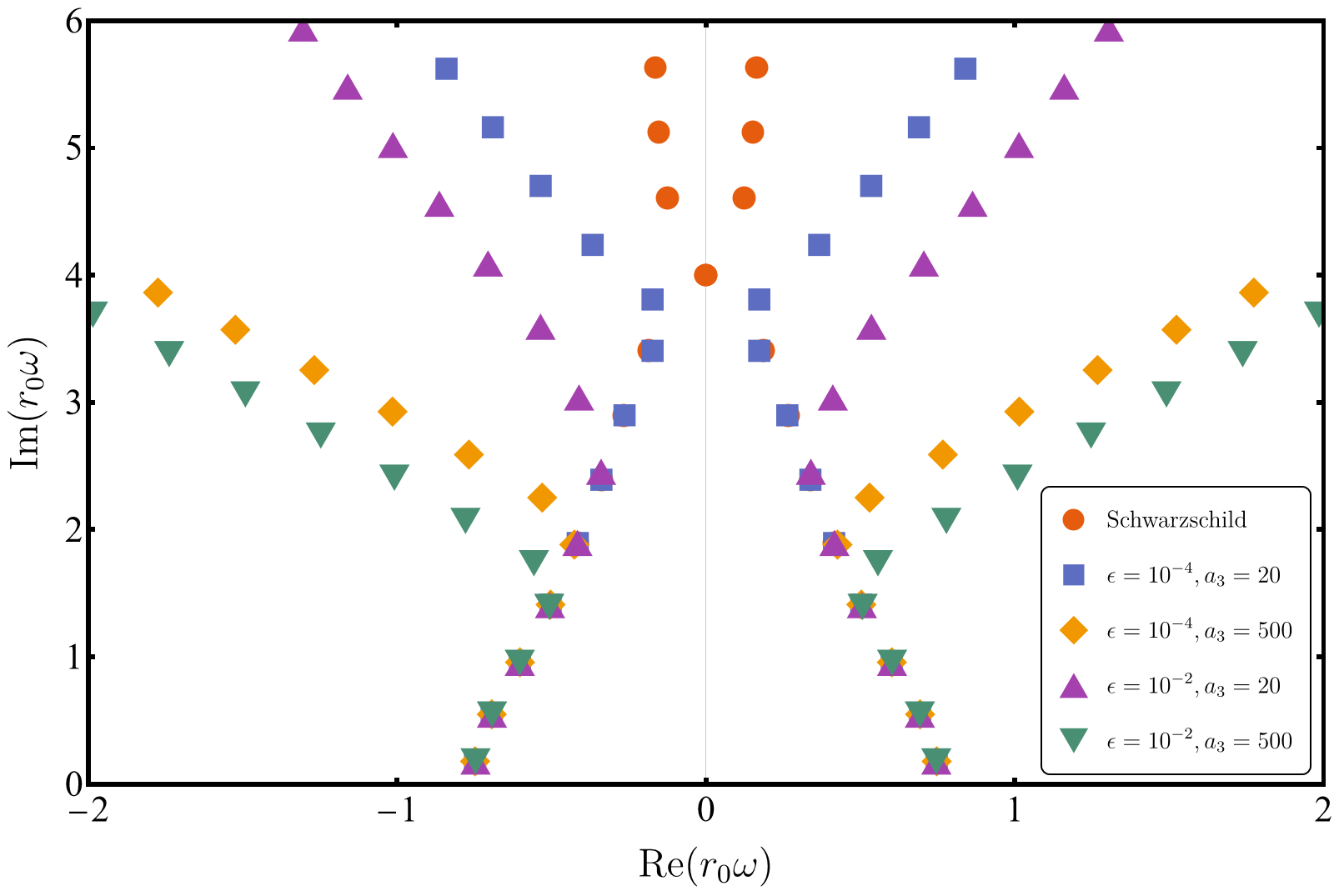}
		\caption{Left Panel: Density distribution for the RZ spacetime parametrized by $a_3$ and $\epsilon$, which introduce small deformations to the Schwarzschild spacetime near the event horizon. The dashed vertical line represents the position of the light ring, where the density profile attains its  maximum when $a_3=1$ (blue line). As $a_3$ grows, the peak moves towards the horizon, settling at the limiting value $r \approx 1.21 \, r_0$ when $a_3 \rightarrow \infty$ (yellow line). The inset shows the relative difference between the deformed $\ell=2$ axial potential against the corresponding Regge-Wheeler potential for Schwarzschild black holes. Right Panel: QNM spectrum associated with $\ell=2$ axial perturbations for different choices of the parameters $\epsilon$ and $a_3$. Instability for high overtones is triggered as $\epsilon$ and $a_3$ vary, even though deviations from the vacuum Schwarzschild spacetime are small. The parameter $a_3$ controls the opening of QNM branches, whereas $\epsilon$ regulates its offset.} 
		\label{fig:density} 	
	\end{figure*}

	\subsection{Physically motivated deformed metric}\label{sec:MetricPar}
    
    Given the {\em ad-hoc} nature of the RZ metric \eqref{eq:krz-metric}, one should not expect all choices of the free deformation parameters to result in physically realistic spacetimes.
    We follow Ref.~\cite{Cardoso:2024mrw} to motivate the configuration
considered in this work.
We first impose conditions determined by the weak-field limit and arising from a consistent asymptotic structure of the spacetime. Observational constraints on post-Newtonian parameters suggest that $|a_0| \lesssim 10^{-4}$ and $|b_0| \lesssim 10^{-4}$ based on solar system measurements~\cite{Will:2014kxa,Rezzolla2014}. Other studies also examine constraints on post-Newtonian coefficients derived from binary pulsars and binary black holes~\cite{Will:2014kxa,LIGOScientific:2016lio}, although establishing a direct relation to the RZ constants lies beyond the scope of this work. For our analysis, we adopt the simplifying assumptions
\begin{equation}
    \label{eq:cond_PN}
    a_0 = \epsilon, \quad b_0 = 0,
\end{equation}
where $\epsilon$ is treated as a free theoretical parameter. Specifically, we consider $\epsilon = 10^{-4}$, consistent with observational constraints, and also explore $\epsilon = 10^{-2}$ to test the stability of QNMs. 
   The choice $b_0=0$ ensures that the mass function asymptotically approaches the ADM mass, i.e., \(m(r) \to M_{\text{ADM}}\) as \(r \to \infty\). 
   
Considering the behavior in the vicinity of the black hole, we also require that the Hawking temperature, the density at the horizon and the pressures at the horizon coincide with the corresponding values for a Schwarzschild spacetime, namely
    \beq T_H = 1/(4 \pi r_0), \quad \rho(r_0)=p_r(r_0) =p_t(r_0)= 0.
    \eeq    
The equations above, together with \eqref{eq:cond_PN}, imply that 
    \beq
	\label{eq:cond_rho_T}
	a_1 = \epsilon, \quad b_{1} = 0, \quad a_2 = -2.
	\eeq
	After imposing \eqref{eq:cond_PN} and \eqref{eq:cond_rho_T}, one is still left with the freedom to choose the parameter $\epsilon$, and all coefficients $a_k$ for $k\geq 3$. The simple choice $a_4 = 0$ reduces the parameter space to a bidimensional family of spacetimes fixed by $\epsilon$ and $a_3$. 
    The energy-momentum tensor of these deformed Schwarzschild spacetimes scales with $\epsilon$, making this parameter a natural measure of the intensity of the deviations triggered by $a_3$. In particular, $a_3\geq 1$ ensures that $\rho(r)>0$ in the exterior black hole region $r\geq r_0$. 
    We highlight that the equation of state for the radial pressure is precisely $p_r (r) = - \rho(r)$, a characteristic commonly associated with  $\Lambda$CDM models that account for dark energy in cosmology. This property stands out as an interesting peculiarity of the spacetime, although a detailed investigation of its significance and implications lies beyond the scope of this work. 
    We also remark that the fluid described by \eqref{stresstensor}-\eqref{stresstensor2} is anisotropic, as the radial and tangential pressures differ. Anisotropic models like this have been widely employed in the literature to describe both stars and black holes~\cite{1975A&A....38...51H,Herrera:1997plx,Mak:2001eb,Harko:2002pxr,Harko:2002db,Herrera:2004xc,Abreu:2007ew,Silva:2014fca,Cho:2017nhx,Visser:2019brz}.  
      
   In Fig.~\ref{fig:density} we review some properties of the RZ spacetime (left panel) together with the QNM spectrum for $\ell=2$ axial gravitational perturbations (right panel). The left panel illustrates the density profile of the spacetime and the associated effective potential for axial perturbations, as a function of the radial coordinate, for both $a_3=1$ and $a_3 \to \infty$. As emphasized by Ref.~\cite{Cardoso:2024mrw}, for $a_3=1$ the density profile reaches its maximum value $8 \pi \rho r_0^2 \approx 6.58 \times 10^{-2}\epsilon$ exactly at the photon sphere $r=3 r_{0}/2$.    
   As $a_3$ increases, the peak shifts closer to the event horizon, ultimately approaching the value $8 \pi \rho r_0^2 \approx 2.82\times 10^{-1}\, \epsilon$ at $r \approx 1.21 \, r_{0}$ in the limit $a_3\rightarrow \infty$. 
    The inset of the left panel of Fig.~\ref{fig:density} depicts the relative difference between the effective potential for axial perturbations and the corresponding Regge-Wheeler expression for Schwarzschild black holes. The maximum difference is of the order of $\sim 10^{-1}\, \epsilon$,  occuring around $r\sim 2 r_0$. One can thus conclude that the RZ spacetime here considered introduces modifications for axial gravitational perturbations close to (but not at) the event horizon of a Schwarzschild black hole.

The right panel of Fig.~\ref{fig:density} exhibits the QNM spectrum associated with $\ell=2$ axial perturbations, following the analysis of \cite{Cardoso:2024mrw}, for four distinct configurations of the RZ black hole:  $(\epsilon,a_3) = (10^{-4}, 20)$, $(10^{-4}, 500)$, $(10^{-2}, 20)$, and $(10^{-2}, 500)$.  The instability of the overtones becomes evident, as the new QNMs differ from their respective Schwarzschild values by orders of magnitude larger than the parameter $\epsilon$. In fact, the parameter $a_3$ controls the opening of new QNM branches, while $\epsilon$ determines the overtone offset where the instability is triggered. More precisely, for fixed $\epsilon$, variations in $a_3$ significantly affect the slope of the overtone sequence, with higher overtones becoming progressively more horizontally aligned as $a_3$ increases. Moreover, increasing $a_3$ accelerates the onset of spectral instabilities. For instance, when $a_3 = 20$, instability emerges at the seventh overtone for $\epsilon = 10^{-2}$ and at the ninth overtone for $\epsilon = 10^{-4}$. In contrast, for $a_3 = 500$, instability arises at the fifth overtone for $\epsilon = 10^{-2}$ and at the sixth overtone for $\epsilon = 10^{-4}$. The methods employed to calculate the spectra shown in Fig.~\ref{fig:density} are reviewed in Sec.~III, along with the tools used in Sec.~IV to compute the associated pseudospectra.

Recognizing that the RZ metric is able to incorporate key elements of a physically reasonable spacetime triggering QNM instabilities from first principles, we will employ it to investigate an intriguing and seemingly paradoxical phenomenon described in Ref.~\cite{PhysRevX.11.031003}: QNM spectra generated by random perturbations to the effective potential are expected to be stable. In other words, we analyze the stability of the QNM spectra of perturbed Schwarzschild black holes, utilizing the RZ metric with the parametrization introduced in this section to model the associated deformations, rather than relying on the agnostic, yet unphysical, approach of random perturbations.

	%%%%%%%%%%%%%%%%%%%%%%%%%%%%%%%%%%%%%%%%%%%%%%%%%%%%%%%%%%%%%%%%%%%%%%%%%%%%%%%%%%%%%%%%%%%%%%%% 
	\section{Pseudoespectrum} 
	\label{Sec:pseudo-krz}
    
By regularizing the QNM eigenfunction at the black hole horizon and at the asymptotic region, the hyperboloidal framework provides a useful approach to study QNM spectra~\cite{Zenginoglu:2011jz,Panosso_Macedo_2024,PanossoMacedo:2024nkw}. Apart from allowing for an efficient computation of QNM frequencies by reformulating the associated eigenvalue problem on a compactified domain, the framework is fundamental for the calculation of the black hole pseudospectrum. This section reviews the theoretical tools employed to study the RZ spacetime and the stability properties of its QNM spectra.

    \subsection{Hyperboloidal Approach} 
	\label{Sec:hyper-krz}

    We start by reformulating the spacetime and the QNM problem discussed in Sec.~\ref{Sec:def_krz} using the hyperboloidal framework. Following the scri-fixing technique \cite{Zenginoglu:2007jw,Zenginoglu:2011jz}, and using the notation of Ref.~\cite{Panosso_Macedo_2024}, we define the hyperboloidal coordinates ($\tau$, $\sigma$)    
   through the transformations
	\begin{equation} \label{hypercoord}
		t = r_0 \left[ \tau - H(\sigma) \right] , \quad  r = \dfrac{r_0}{\sigma},
	\end{equation}
    where $(t, r)$ are the original temporal and radial coordinates, and $H(\sigma)$ is known as the height function~\cite{Panosso_Macedo_2024}.
	In the equation above, the radial compactification assumes the so-called minimal gauge~\cite{Panosso_Macedo_2024}. The compact radial coordinate is defined in the domain $\sigma\in[0, 1]$, with the future null infinity corresponding to $\sigma = 0$ and the event horizon to $\sigma = 1$. It is also convenient to identify the dimensionless tortoise coordinate $x(\sigma)$, related to the standard tortoise coordinate $r_*$, defined in \eqref{eq:def_r_*}, via
    \begin{equation}
    {\rm x}(\sigma) = \dfrac{r_*[r(\sigma)]}{\lambda} = {\rm x}_{\infty}(\sigma) +  {\rm x}_{r_0}(\sigma)   +  {\rm x}_{\rm res}(\sigma). \label{eq:x_decomposition}
    \end{equation}

    Eq.~\eqref{eq:x_decomposition} identifies the quantities ${\rm x}_{\infty}(\sigma)$ and ${\rm x}_{r_0}(\sigma)$, which exhibit singular behavior at the future null infinity and at the black hole event horizon, respectively. The residual term ${\rm x}_{\rm reg}(\sigma)$ accounts for additional terms that are regular in the exterior black hole region $\sigma\in[0,1]$. Since the expressions for ${\rm x}_{\infty}(\sigma)$, $x_{r_0}(\sigma)$ and ${\rm x}_{\rm res(\sigma)}$ are quite lengthy and offer limited insight, we refrain from presenting them explicitly here. Ref.~\cite{Panosso_Macedo_2024} provides the steps for their calculation, which leads to the following height function in the minimal gauge, within the in-out strategy~\cite{Panosso_Macedo_2024}
    \begin{equation}
		H (\sigma) = -{\rm x}_{\infty}(\sigma) + {\rm x}_{r_0}(\sigma) + {\rm x}_{\rm reg}(\sigma).
	\end{equation}

    The coordinate transformation \eqref{hypercoord} yields a conformal rescaling of the line element \eqref{eq:krz-metric} as $ds^{2} = \Omega^{-2} d\bar{s}^{2}$, with $\Omega = \sigma/r_0$. The conformal metric is 
    \begin{equation} \label{conformalmetric}
		d\bar{s}^{2} = \ -p(\sigma) d\tau^{2} + 2 \gamma(\sigma) d\tau d\sigma\ + w(\sigma) d\sigma^{2} + d\varpi^{2},   	
  \end{equation}
  with $d\varpi^{2} = d \theta^{2} + \sin^{2} \theta d \phi^{2}$ and the metric components given by the following  functions,
  \bea \label{reg_functionsI}
    p(\sigma) &=& -\dfrac{1}{x'(\sigma)}, \\  \label{reg_functionsII}
    \gamma(\sigma) &=& p(\sigma) H'(\sigma), \\  \label{reg_functionsIII}
    w(\sigma) &=& \dfrac{1-\gamma(\sigma)^2}{p(\sigma)}.
  \eea
We remark that the above functions are regular in the entire exterior domain $\sigma \in [0, 1]$. The hyperboloidal coordinate change \eqref{hypercoord} also induces a transformation of the function $\psi(r)$ in \eqref{eq:QNM_operator_SchwarCoord}, namely
  \beq
    \psi(r) = Z(\sigma) \bar \psi(\sigma), \quad Z(\sigma) = e^{s H(\sigma)},
  \eeq
  with $s = - i r_0 \omega$ defining the dimensionless frequency. The function $Z(\sigma)$ geometrically incorporates the outgoing behavior of QNMs [recall the boundary conditions \eqref{eq:BC}] through the height function $H(\sigma)$~\cite{Zenginoglu:2011jz,Panosso_Macedo_2024}.  
  
  With $\bar \phi = s \bar \psi$, the wave equation \eqref{eq:QNM_operator_SchwarCoord} takes the explicit form of an eigenvalue problem,
  	\begin{equation}
   \label{eq:QNM_operator_Hyp}
   L\, \boldsymbol{u}= s \, \boldsymbol{u}, \quad
		L =  \left(
		\begin{array}{cc}
			0 & 1\\
			L_{1} & L_{2}
		\end{array}
		\right),
  \quad
  \boldsymbol{u} = \left( \begin{array}{c}
			\bar \psi(\sigma) \\
			\bar \phi(\sigma)
		\end{array}
  \right),
	\end{equation}	
 	where the matricial operator $L$ depends on the linear operators $L_1$ and $L_2$ defined by
    \begin{subequations} 
    \bea
		L_{1} =& \dfrac{1}{w(\sigma)} \bigg[ \partial_{\sigma} (p(\sigma) \partial_{\sigma}) - q_\ell(\sigma) \bigg], \\
		L_{2} =& \dfrac{1}{w(\sigma)} \bigg[ 2 \gamma(\sigma) \partial_{\sigma} + \partial_{\sigma} \gamma(\sigma) \bigg],
	\eea
    \end{subequations}
with the conformal potential $q_{_\ell}$ given by 
	\begin{equation}
    \label{eq:conf_pot}
		q_\ell(\sigma) = \dfrac{r_{0}^{2}}{p(\sigma)}V_{_\ell}(r).
	\end{equation}
 We note that the operator $L$ is non-self-adjoint~\cite{PhysRevX.11.031003} and that such a characteristic may indicate spectral instability, i.e., small perturbations of this operator could trigger large deviations of the QNM spectrum, when compared to the unperturbed case. This phenomenon can be qualitatively analyzed through the pseudospectrum of the operator, which we define in the next section.

    \subsection{Pseudospectrum}\label{sec:pseudospectrum}
With the equation that defines QNMs reformulated as an eigenvalue problem in \eqref{eq:QNM_operator_Hyp}, we can apply tools from the theory of non-normal operators~\cite{trefethen2005spectra,Davie07,Sjostrand2019} to examine the resolvent ${\cal R}_L(s) = \left(s \, {\rm Id}-L\right)^{-1}$ beyond its QNM spectrum (here, ${\rm Id}$ denotes the identity operator). Specifically, the spectrum of $L$ is determined by the points $s_n$ for which ${\cal R}_L(s_n)$ is singular. The $\varepsilon$-pseudospectrum $\varsigma^\varepsilon(L)$ expands this notion into
    \beq
    \label{eq:pseudospectrum_def}
    \varsigma^\varepsilon(L) = \left\{ s\in {\mathbb C} \,\, : \,\, \left|\left| {\cal R}_L(s) \right|\right|^{-1} < \varepsilon \right\},
    \eeq    
    which formally captures the sensitivity of the operator to perturbations, allowing a non-modal analysis of the system~\cite{PhysRevX.11.031003,Jaramillo:2022kuv}. The limit $\varepsilon \rightarrow 0$ recovers the spectrum of the operator $L$, denoted by $\varsigma(L)$. 

    We emphasize that the definition of the pseudospectrum depends on the choice of the underlying scalar product~\cite{PhysRevX.11.031003,Gasperin:2021kfv,Besson:2024adi}, which we take to be the energy associated with the field satisfying the wave equation. Thus, the energy norm associated with the state vector ${\boldsymbol u}$ reads
    \begin{align}\label{energy norm}
\hspace{-0.25cm} ||{\boldsymbol u}||^2_{_{E}}  
=\frac{1}{2}\int_0^1 \! \! \left[w(\sigma)|\bar \phi|^2
+ p(\sigma)|\partial_\sigma\bar\psi|^2 + q(\sigma) |\bar\psi|^2\right] \! d\sigma.
\end{align}
    From the above expression, one derives the energy scalar product between two state vectors ${\boldsymbol u}_1$ and ${\boldsymbol u}_2$ as follows, 
    \bea
    \label{eq:energy_scalar_prod}
    \braket{{\boldsymbol u}_1|{\boldsymbol u}_2}_{_{E}} &=& \frac{1}{2}\int_0^1 \! \! \left[w(\sigma) \bar \phi_1^\ast \bar \phi_2
+ p(\sigma)\partial_\sigma \bar\psi_1^\ast \partial_\sigma\bar\psi_2 \right.\nn  \\ &&
\left. \qquad + q(\sigma) \bar\psi_1^\ast \bar\psi_2 \right] \! d\sigma,
    \eea
    with the $\ast$ operation corresponding to complex conjugation. The energy scalar product~\eqref{eq:energy_scalar_prod} directly induces a norm applicable to the operator appearing in \eqref{eq:pseudospectrum_def}, which we discuss below using a discretization scheme based on Chebyshev spectral methods --- see e.g.~\cite{PhysRevX.11.031003} for more details.

\subsection{Numerical Approximation}
\label{NumericalApprox}
	 We discretize the system in terms of a Chebyshev collocation point spectral method \cite{PhysRevX.11.031003}. We fix a truncation parameter $N \in \mathbb{N}$ and introduce Chebyshev-Lobatto grids points
    \beq \label{xi}
    \chi_i = \cos\left(\dfrac{\pi i}{N} \right), \quad i = 0,1, \, \cdots, N,
    \eeq
    which discretize the Chebyshev polynomial's domain of dependence $\chi \in[-1,1]$. In terms of this discretization, the derivative and integration operators assume discrete matricial forms $\hat D_{\chi}$ and $\hat {\cal G}_{\chi}$, respectively
    \beq
    \partial_\chi \rightarrow \hat D_{\chi}, \quad \int_{-1}^{1}d\chi \rightarrow \hat {\cal G}_{\chi}.
    \eeq
    The components $\hat D_{\chi}$ and ${\cal G}_{\chi}$ are given explicitly in the appendix of Ref.~\cite{PhysRevX.11.031003}. 
    
    To map the spectral coordinates $\chi\in[-1,1]$ into the physical domain $\sigma\in[0,1]$, we define the transformation
    \begin{subequations}
    \label{eq:y_chi}
    \bea
    \sigma(\chi) &=& \dfrac{1 + y(\chi) }{2},\\
    y(\chi) & =& 1 - \dfrac{2 \sinh\left[ \kappa \left( 1 - \chi \right) \right]}{\sinh(2\kappa)},
    \eea
    \end{subequations}
    which is based on the so-called analytical mesh-refinement (AnMR)~\cite{PanossoMacedo:2022fdi} and depends on the free parameter $\kappa$. For $\kappa = 0$, one recovers the identity $y(\chi)=\chi$. For $\kappa >0$, the function $y(\chi)$ still maps the interval $[-1,1]$ into itself, but accumulates the grid points around the boundary $y=1$ of the domain (which corresponds to $\sigma =1$). Consequently, it resolves more accurately eventual strong gradients developing around the horizon in the conformal potential $\bar V_\ell(\sigma)$ given by \eqref{eq:conf_pot}. A more detailed account of AnMR and its application to the QNM problem using the hyperboloidal framework will be discussed elsewhere~\cite{Yi2025}.

    In terms of the AnMR mapping \eqref{eq:y_chi}, the state vector ${\boldsymbol u}$ is discretized as $\vec u$. To represent the discrete differentiation and integration operators along the coordinate $\sigma$, respectively $\hat D_{\sigma}$ and $\hat{\cal G}_{\sigma}$, we employ element-wise (Hadamard) products, denoted by the circle ($\circ$) operator. Thus, for a given vector $\vec{v}$ and matrix $\hat M$, the element-wise product reads 
    \beq
    \left(\vec{v}\circ \hat M\right)_{ij} = v_i \, M_{ij}.
    \eeq
    Besides, for a given function $f(\sigma)$, we abuse the vector notation to represent the discrete form of $1/f(\sigma_i)$ via $\vec{f}^{-1}$, with components
    \beq
    \left(\vec{f}^{-1}\right)_i = \left(f_i\right)^{-1}.
    \eeq
    With the above notation, the differentiation and integration operators read
    \beq
    \label{eq:disc_derv}
    \hat D_{\sigma} = \vec{J}^{-1} \circ \hat D_{\chi}, \quad \hat{\cal G}_{\sigma} = \vec J \circ \hat{\cal G}_{\chi}, 
    \eeq
    with $\vec J $ defined by its components through
    \beq
    J_i = \dfrac{d \sigma(\chi_i)}{d\chi}.
    \eeq
    
    The evaluation of the functions in \eqref{reg_functionsI}-\eqref{reg_functionsIII} and \eqref{eq:conf_pot} at the grid points $\sigma_i = \sigma(\chi_i)$ yields the vectors $\vec{p}$, $\vec{p'}$, $\vec{\gamma}$, $\vec{\gamma'}$, $\vec{w}$ and $\vec{q}_{\ell}$, from which we construct the discrete operator
    \begin{subequations} \label{eq:discretized_L}
    \bea
    \hat{L}&=& \left( 
    \begin{array}{cc}
    \mathbb{0} & \mathbb{1} \\
    \hat{L}_1 & \hat{L}_2
    \end{array}
    \right), \\
    \hat{L}_1 &=& \vec{w}^{-1}\circ \left( \vec{p}\circ \hat D^2_{\sigma} +\vec{p'}\circ \hat D_{\sigma} - \vec{q}_{\ell}\circ \mathbb{1}  \right), \\
    \hat{L}_2 &=& \vec{w}^{-1}\circ \left( 2\vec{\gamma}\circ \hat D_{\sigma} +\vec{\gamma'}\circ \mathbb{1}  \right),
    \eea
    \end{subequations}
    where $\mathbb{0}$ and $\mathbb{1}$ are, respectively, null and identity matrices of size $(N+1)\times (N+1)$.
    Moreover, the calculation of the energy norm  \eqref{energy norm} follows from
    \begin{subequations}
     \bea
	 && ||\vec u||^2_{_{E}}  = \vec{u}^\ast \cdot \hat G \cdot \hat u,  \\
     && \hat G = \left(
        \begin{array}{c|c}
      \hat D_{\sigma}^t \cdot ( \vec p  \circ \hat{\cal G}_{\sigma}  ) \cdot \hat D_{\sigma} +  \vec q \circ \hat{\cal G}_{\sigma}    & \mathbb{0} \\
      \hline
           \mathbb{0}  & \vec w \circ \hat{\cal G}_{\sigma}
        \end{array}
        \right),
     \eea
     \end{subequations}
    where the dot ($\cdot$) represents the usual matrix vector multiplication.  
       
	Finally, the $\varepsilon$-pseudoespectrum $\varsigma^{\varepsilon} (\hat L)$ is given by
	\begin{equation} \label{eq:pseudospectrum_def2}
		\varsigma^{\varepsilon} (\hat L) = \{ s \in \mathbb{C}: {\mathfrak s}^{\text{min}} \left( \hat A \right) < \varepsilon \}, \quad \hat A = \lambda \mathbb{I} - \hat L,
	\end{equation}
    with $\mathbb{I}$ the identity matrix of size $2(N+1)\times 2(N+1)$,
	and ${\mathfrak s}^{\text{min}}$ the smallest of the generalized singular values, i.e.,
	\begin{equation}
		{\mathfrak s}^{\text{min}}(\hat A) = \text{min} \left\{\sqrt{s} : s \in \varsigma \left(\hat A^{\dagger}\hat A\right)\right\} .
	\end{equation}
    Note that the calculation of ${\mathfrak s}^{\text{min}}$ requires the discrete adjoint $\hat{L}^{\dagger}$ of the operator $\hat L$ with respect to the energy scalar product, given by
	\begin{equation}
		\hat L^{\dagger} = \left(\hat G\right)^{-1} \cdot \left(\hat L^{*}\right)^t \cdot \hat G.
	\end{equation}

  \begin{figure*}[!htpb]
		\centering
		\includegraphics[width = 1 \linewidth]{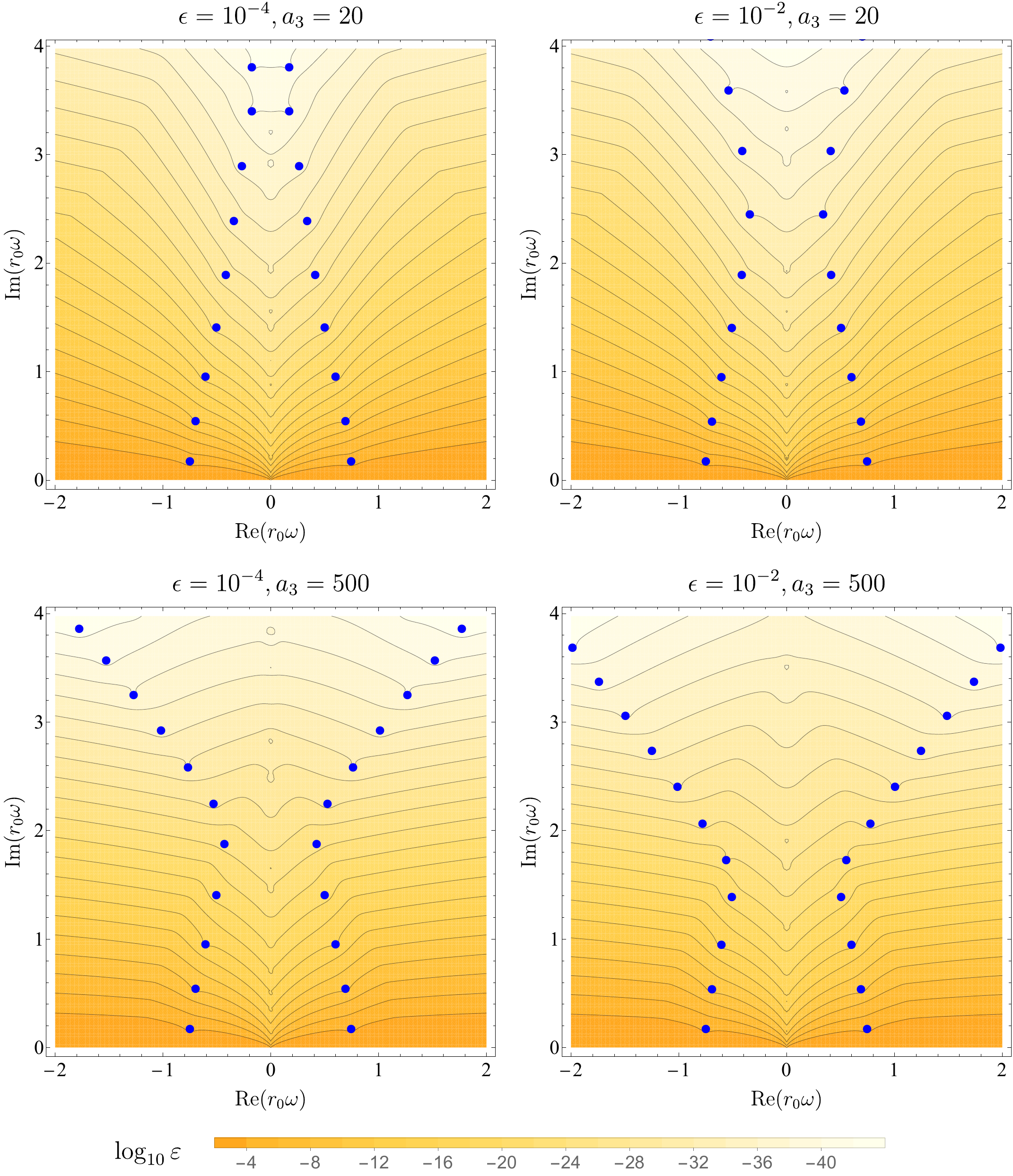}
		\caption{The $\varepsilon$-pseudospectrum, defined in \eqref{eq:pseudospectrum_def} and \eqref{eq:pseudospectrum_def2}, for four distinct configurations of the RZ spacetime. The top panels correspond to $a_{3} = 20$ with $\epsilon = 10^{-4}$ (left) and $\epsilon = 10^{-2}$ (right), while the bottom panels represent $a_{3} = 500$ with $\epsilon = 10^{-4}$ (left) and $\epsilon = 10^{-2}$ (right). The color scale in the background of each panel shows $\log_{10} \varepsilon$, and the black contour lines indicate the level sets of constant $\varepsilon$. In all panels, the contour lines are not bounded in the plot range, spreading across the complex plane  and signaling spectral instability (as opposed to closed contour lines around the eigenvalues, which would indicate spectral stability).
        Note that, even around QNMs which have already been significantly displaced from their Schwarzschild counterparts (see the right panel of Fig.~\ref{fig:density}), the contour lines of the pseudospectra indicate instability with respect to additional perturbations.
        This observation contrasts with Ref.~\cite{PhysRevX.11.031003}, which suggests stability if the spectra had resulted form random perturbations.
        }
		\label{fig:todos_plots} 	
	\end{figure*}

	\section{Results}
  In this section, we investigate the stability of the QNM spectra for the RZ spacetime defined in Sec.~\ref{sec:MetricPar}. Specifically, Sec.~\ref{subsec:pseudo} presents a comprehensive analysis of the $\varepsilon$-pseudospectrum associated with the RZ spacetime. Subsequently, in Sec.~\ref{subsec:unstable}, we consider an additional ({\em ad-hoc}) perturbation to the effective potential of the RZ spacetime to further explore the spectral instabilities predicted by the pseudospectra.
    
    The numerical analysis requires fixing the discretization parameter $N$, and  the mesh-refinement parameter $\kappa$, see Eqs.~\eqref{xi} and \eqref{eq:y_chi}. The choice of the truncation parameter $N$ is based on the following procedure. For each set of parameters $(\epsilon, a_3)$, we compute the QNMs of the RZ spacetime with $N = 100$. We then incrementally increase $N$ in steps of $50$ until the relative difference between the last and second-to-last values of the absolute magnitude of the eighth overtone become smaller than the specified threshold $\Delta = 10^{-2}$. For all combinations of $\epsilon$ and $a_3$ analyzed in our work, we found that $N = 300$ satisfied the criterion. Consequently, this value was used to generate the results presented in this article, including the pseudospectra.

To optimize the AnMR procedure, we considered a range of values for $\kappa$ spanning the interval $[0, 10]$, with increments of $0.1$. For each spacetime configuration and each $\kappa$, we computed the Chebyshev-Lobatto coefficients $c_k$ corresponding to the discrete representation of the conformal potential $\tilde{V}_{l}(\sigma)$ defined in \eqref{eq:conf_pot}. The optimal value of $\kappa$ is defined as the value which minimizes the magnitude of the smallest Chebyshev-Lobatto coefficient. We have found that this optimal value depends on the metric parameter $a_3$, and for the configurations employed in our analyzes, i.e., $a_3 = 20$ and $a_3 = 500$, we found that the numerical accuracy was optimized for $\kappa = 2.2$ and $\kappa = 3.8$, respectively.

 \subsection{Pseudoespectrum}
 \label{subsec:pseudo}
We now analyze the pseudospectra associated with the wave operator for axial modes in the RZ spacetime. Focusing on the quadrupole modes ($\ell=2$) of the RZ parametrization described in Sec.~\ref{sec:MetricPar}, the four panels of Fig.~\ref{fig:todos_plots} display the pseudospectra corresponding to the following configurations: $(\epsilon, a_3) = (10^{-4}, 20)$ (top left), $(\epsilon, a_3) = (10^{-2}, 20)$ (top right), $(\epsilon, a_3) = (10^{-2}, 500)$ (bottom left), and $(\epsilon, a_3) = (10^{-2}, 500)$ (bottom right). The blue points in Fig.~\ref{fig:todos_plots} correspond to the QNMs shown in the right-hand panel of Fig.~\ref{fig:density}. These QNMs are computed as the eigenvalues $s$ of the operator $L$ defined in \eqref{eq:QNM_operator_Hyp}, according to the hyperboloidal approach described in Sec.~\ref{sec:pseudospectrum}. In practice, we use the discretized forms of $L$ and  $\boldsymbol{u}$, obtained from the discretization of the coordinate $\sigma$ as in \eqref{eq:y_chi}, with $L$ given explicitly by \eqref{eq:discretized_L}.

Recall that the QNMs of the RZ metric are already destabilized relative to their Schwarzschild counterparts, with the parameter $a_3$ governing the opening of new QNM branches and $\epsilon$ determining the overtone offset where the instability is triggered. Motivated by the arguments in Ref.~\cite{PhysRevX.11.031003}, we hypothesized that a perturbed QNM spectrum (such as the QNM spectra for the RZ spacetime) may remain stable under additional perturbations.
However, our findings do not support this hypothesis. On the contrary, in all cases depicted in Fig.~\ref{fig:todos_plots} the pseudospectrum analysis reveals spectral instability for the underlying QNMs. Specifically, we observe the same spreading pattern in the pseudospectrum level sets as in other black hole spacetimes, with the contour lines opening up across the complex plane.

This result may have been expected for the configuration $(\epsilon, a_3) = (10^{-4}, 20)$ (top left in Fig.~\ref{fig:todos_plots}), for which the QNMs of the  RZ black hole do not deviate significantly from the Schwarzschild values in the region of the complex plane considered. Nevertheless, even as $a_3$ increases (thus triggering the QNM instability associated with the Schwarzschild-like metric), the pseudospectrum retains the same qualitative behavior, showing no indication of increased stability.
This effect is particularly evident for the configurations with $a_3 = 500$ (bottom panels in Fig.~\ref{fig:todos_plots}). In this case, as seen in the right panel of Fig.~\ref{fig:density}, the QNMs already deviate significantly from their Schwarzschild values for overtones with $n \geq 4$.
Despite that, there is no indication in the pseudospectra that these new QNMs (i.e., the QNMs of the RZ black hole) are stable,  as we do not observe the characteristic signature of spectral stability in that region --- namely, a flat pseudospectrum with concentric circles around the QNMs. The next section explores the effects of the diagnosed spectral instability.

\subsection{Perturbing the unstable}
\begin{figure}[!b]
		\centering
		\includegraphics[width =\columnwidth]{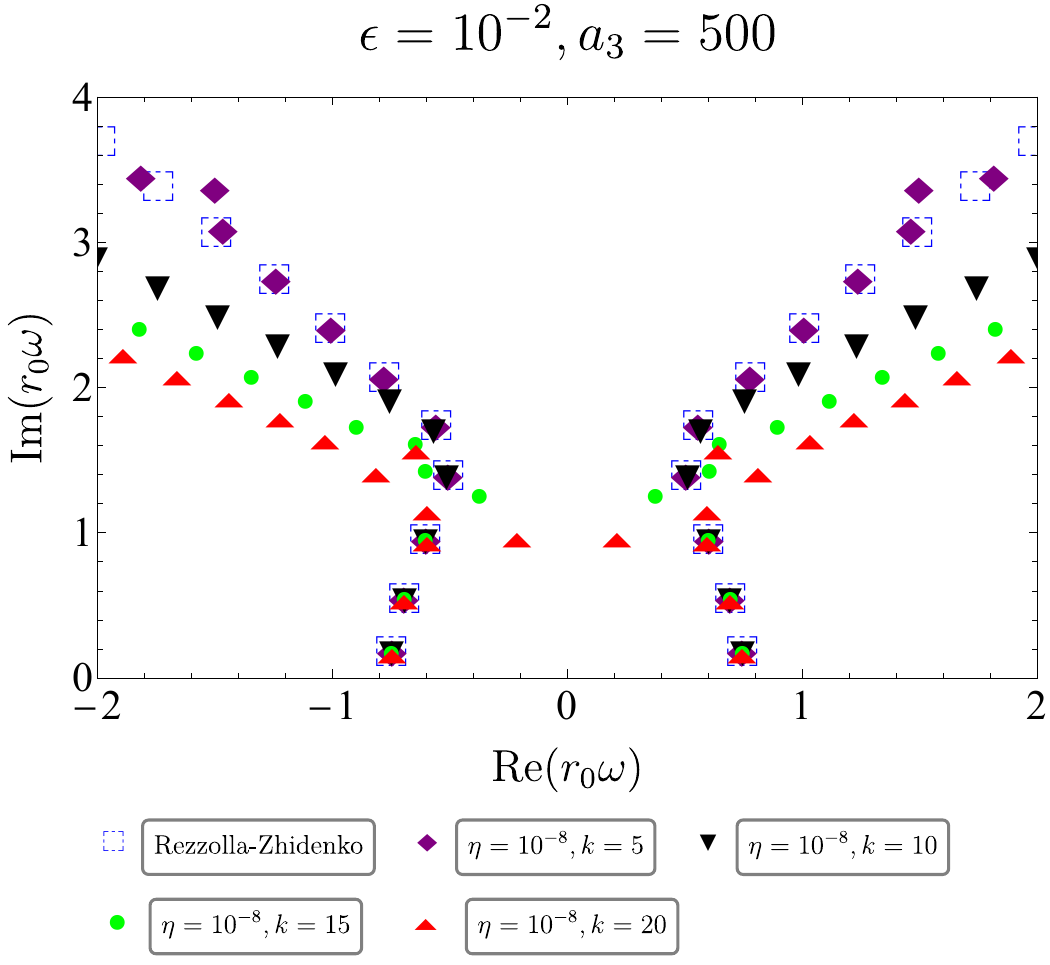}
		\caption{QNM spectra of a Schwarzschild-like black hole defined by the RZ parameters $\epsilon = 10^{-2}$ and $a_{3}$ = 500. The square (blue) markers correspond to the QNMs of the RZ spacetime without additional deformations. The other marks, which refer to the QNM spectra in the presence of a sinusoidal deformation of the form \eqref{eq:delta_q}, with $\eta = 10^{-8}$ and $k = 5$, $k = 10$, $k = 15$ and $k = 20$, reveal the spectral instability of the RZ spacetime. Note that perturbations with low $k$ values cause minor shifts in the QNMs, but as $k$ increases, spectral instability is  exacerbated, demonstrating that multiple competing effects [in this case, deformation from the RZ parametrization and from the sinusoidal term \eqref{eq:delta_q}] contribute to the overall instability of the modes.}     
		\label{fig:epsilon_a3} 	
	\end{figure}

    	\begin{figure*}[!htpb]
		\centering
		\includegraphics[width = 1 \linewidth]{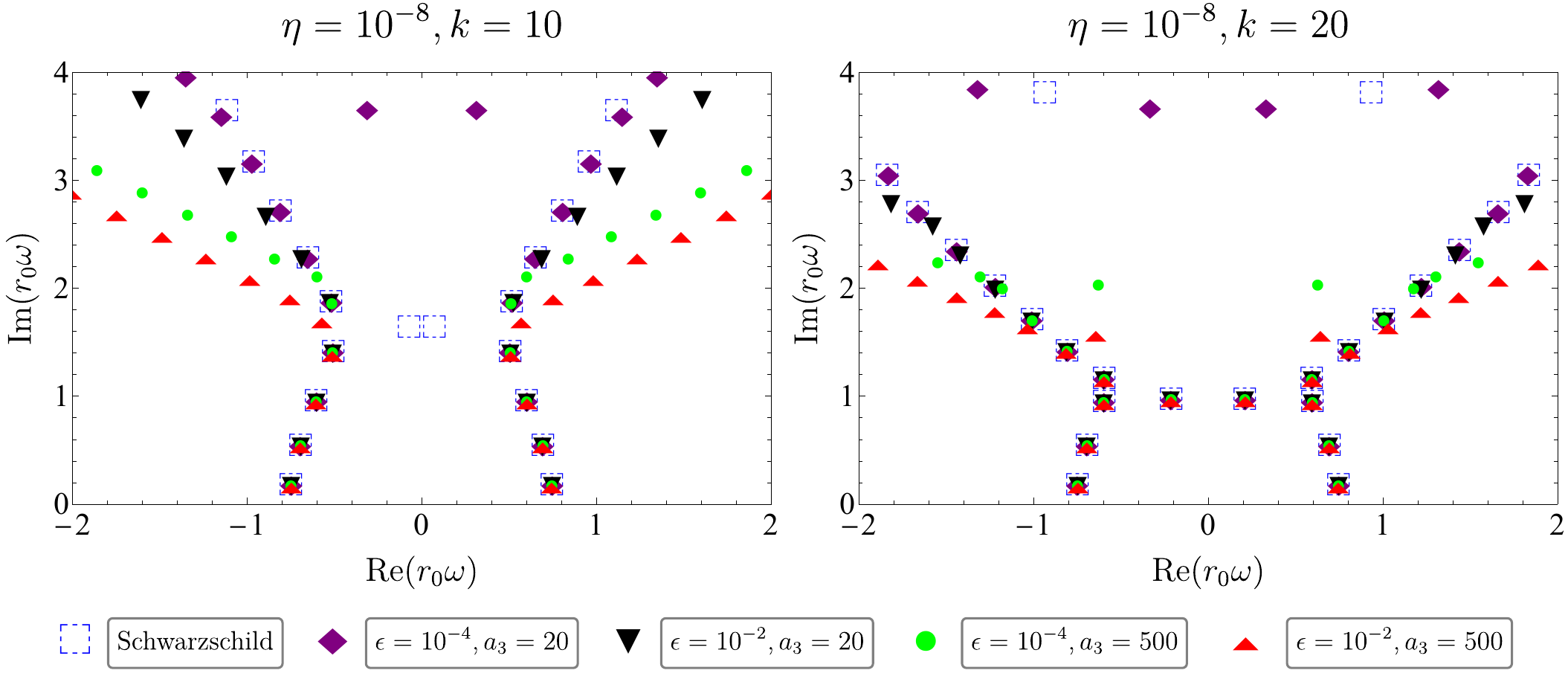}
         \caption{QNM spectra for the Schwarzschild and RZ spacetimes, with associated black hole potentials perturbed by the same sinusoidal source \eqref{eq:delta_q}. The perturbations are defined by $(\eta, k) = (10^{-8}, 10)$ in the left panel and $(\eta, k) = (10^{-8}, 20)$ in the right panel. Markers denote the distinct background metrics: Schwarzschild and deformed Schwarzschild black holes within the RZ framework, with parameters ($\epsilon$, $a_{3}$) =  ($10^{-4}$, $20$), ($10^{-2}$, $20$), ($10^{-4}$, $500$), and ($10^{-2}$, $500$). Overlaps in the spectra indicate that the presence of the sinusoidal source may hinder using overtone instabilities to identify the RZ deformation parameters of a Schwarzschild-like black hole.}
		\label{fig:eps_a3} 	
	\end{figure*}

 \label{subsec:unstable}

To complement the pseudopectrum analyses of Sec.~\ref{subsec:pseudo} and explore the unstable nature of the QNM spectra of the RZ spacetime, we introduce an explicit perturbation into the wave operator \eqref{eq:QNM_operator_Hyp}. 
Specifically, we modify the conformal  potential \eqref{eq:conf_pot} through the transformation $q_\ell \rightarrow q_\ell + \delta q$, incorporating a sinusoidal deformation of the form
\begin{equation}
    \label{eq:delta_q}
    \delta q = \eta \sin (2 \pi k \sigma),
\end{equation}
where $\eta$ represents a small amplitude and $k$ denotes the wavenumber of the perturbation. This {\em ad-hoc} modification of the potential leads to an energy density that oscillates
around the vacuum and may assume negative values. Therefore, it cannot be associated with a realistic astrophysical disturbance at the classical level~\cite{Cardoso:2024mrw}; instead, its origin must be rooted in effective quantum vacuum fluctuations. Nevertheless, as a proof-of-principle, perturbations in the potential of the form \eqref{eq:delta_q} are very effective to probe the spectral instability detected via the pseudospectra analysis~\cite{PhysRevX.11.031003}.

Fig.~\ref{fig:epsilon_a3} illustrates the effect of sinusoidal perturbations applied to the effective potential of the RZ spacetime characterized by the parameters $(\epsilon, a_3) = (10^{-2}, 500)$. The amplitude of the sinusoidal deformation is fixed at a small value, $\eta = 10^{-8}$, and the impact of varying the wavenumber on the QNM spectrum is analyzed for $k = 5$ (purple diamond), $k = 10$ (black downward triangle), $k = 15$ (green circle), and $k = 20$ (red  upward triangle). The QNM spectrum of the RZ black hole without the additional sinusoidal perturbation is plotted in blue squares. We note in Fig.~\ref{fig:epsilon_a3} the same characteristic observed in previous studies (e.g.,~\cite{PhysRevX.11.031003}): sinusoidal perturbations with a relatively low wavenumber do not trigger spectral instability. For example, the QNMs for $(\eta, k) = (10^{-8}, 5)$ show minimal deviation with respect to the original QNMs of the RZ black hole, indicating that the observed behavior is primarily driven by the instability initially introduced by the RZ deformations of the Schwarzschild-like metric. However, perturbations with moderate wavenumbers are sufficient to destabilize the QNMs, amplifying the spectral instability intrinsic to the RZ spacetime.

Fig.~\ref{fig:epsilon_a3} also reveals that QNM instabilities with respect to a vacuum spacetime (Schwarzschild, in this case) may arise from two (or more) competing effects. For instance, the deformation from the RZ parametrization scheme may model an astrophysical effect due to the anisotropic character of the energy momentum tensor~\cite{1975A&A....38...51H,Herrera:1997plx,Mak:2001eb,Harko:2002pxr,Harko:2002db,Herrera:2004xc,Abreu:2007ew,Silva:2014fca,Cho:2017nhx,Visser:2019brz}, whereas the sinusoidal perturbation, as stated before, might be interpreted as a toy-model for an effective quantum vacuum fluctuation. 
This raises the natural question of whether observing overtone instabilities from various sources can provide sufficient information to distinguish the underlying causes of these perturbations. In our specific model, this question translates to determining whether the presence of a sinusoidal perturbation affects our ability to differentiate between a pure Schwarzschild spacetime and a RZ spacetime, which is already perturbed relative to Schwarzschild.

To explore this issue, in Fig.~\ref{fig:eps_a3} we fix the sinusoidal perturbation to the effective potential and compare the QNM spectra for different configurations of the RZ metric. In the left panel, where we fix $(\eta, k) = (10^{-8}, 10)$, there is a clear distinction between the different QNM spectra when the higher overtones are taken into account. 
Observe, however, the qualitative degeneracy between almost every QNM of the Schwarzschild black hole and the corresponding RZ mode when $(\epsilon, a_3) = (10^{-4}, 20)$. This result is expected, as such a choice of RZ parameters trigger the QNM instability only for overtones beyond the range depicted in Fig.~\ref{fig:eps_a3}, as shown in the right panel of Fig.~\ref{fig:density}. Differences between the two configurations are apparent, at most, in terms of modes located internally to new branches, whose theoretical origin and detectability is still an open challenge~\cite{Jaramillo:2021tmt}. On the other hand, for $(\eta, k) = (10^{-8}, 20)$ (right panel of Fig.~\ref{fig:eps_a3}), we observe a higher degree of overlap in the QNM spectra for the Schwarzschild metric and the RZ configurations $(\epsilon, a_3) = (10^{-4}, 20)$, $(10^{-2}, 20)$, and, to some extent, also $(\epsilon, a_3) = (10^{-4}, 500)$. These overlaps suggest that distinguishing the source of the perturbation based solely on the unstable characteristics of the QNM spectra may be challenging.

	%%%%%%%%%%%%%%%%%%%%%%%%%%%%%%%%%%%%%%%%%%%%%%%%%%%%%%%%%%%%%%%%%%%%%%%%%%%%%%%%%%%%%%%%%%%%%%%% 
	
	\section{Final remarks}

    In this work, we analyzed stability and pseudoespectrum regularity of Schwarzschild-like spacetimes within the RZ framework, examining how the deformation parameters in the metric influence the QNM spectrum and the $\varepsilon$-pseudospectrum. We have focused on the physically motivated parametrization of the RZ metric introduced in Ref.~\cite{Cardoso:2024mrw}, which we reviewed in more detail here. Contrary to the assertions in Ref.~\cite{PhysRevX.11.031003}, suggesting that the pseudospectrum becomes more regular in the presence of random perturbations,
   we do not find any sign of improved regularization of the underlying resolvent when the associated wave operator for axial perturbations is derived from a physical Schwarzschild-like spacetime.

    More specifically, by employing the hyperboloidal framework for QNM calculations, together with spectral methods, we investigated how parameters that quantify deviations from the Schwarzschild spacetime in the RZ scheme affect the onset of spectral instabilities and the emergence of patterns in the $\varepsilon$-pseudoespectrum. Within the model from Ref.~\cite{Cardoso:2024mrw}, the Schwarzschild-like spacetimes reduces to a two-dimensional family of solutions, parametrized by $a_3$ and $\epsilon$.  Even though the parameter $a_3$ already destabilizes higher QNMs ovetones from Schwarzschild significantly,  the $\varepsilon$-pseudospectrum associated with the wave equation in the RZ spacetime still diagnosis spectral instability for larger values of $a_3$.  
    Moreover, {\em a posteriori} analyses involving sinusoidal perturbations to the effective potential revealed that these perturbations exacerbate spectral distortions, further underscoring the sensitivity of the spectrum to deviations of the spacetime from Schwarzschild. 

    Although our analysis focused on the axial gravitational modes with $\ell = 2$, preliminary calculations for higher multipoles indicate that the qualitative features of the pseudospectrum and the response to additional perturbations remain consistent with the $\ell = 2$ case. This suggests that our conclusions regarding spectral instability  are robust under variations of the orbital number. Nevertheless, a more detailed analysis involving higher multipoles, and the possibility of simultaneously analyzing multiple $\ell$ modes, is a possible direction for future work.

Future investigations could also expand our analysis by incorporating additional parameters within the RZ framework to describe more general deviations from the Schwarzschild spacetime or to explore the spectral stability of rotating black hole spacetimes beyond the Kerr solution. Besides, evaluating the detectability of overtone instabilities in gravitational wave signals remains a significant challenge~\cite{Spieksma:2024voy}. Progress in this area may benefit from emerging methods for Bayesian analysis tailored to the RZ metric context~\cite{Albuquerque:2025eny}.
Additionally, if the observation of QNM overtone instabilities becomes feasible, our findings indicate that pinpointing the specific origin of the instability could prove difficult, particularly when multiple perturbation sources --- such as environmental effects and deviations arising from alternative theories of gravity --- are simultaneously at play.
\vspace{0.5cm}

\section*{Data Availability}
The data that support the findings of this article are openly available~\cite{data}.
    
	\acknowledgments
    
	PHCS and LTP would like to thank the Strong
	Group at the Niels Bohr Institute (NBI) for their kind hospitality during the final stages of this work.
	The Tycho supercomputer hosted at the SCIENCE HPC center at the University of Copenhagen was used for supporting this work.
    This study was financed in part by the Coordena\c{c}\~ao de Aperfei\c{c}oamento de Pessoal de N\'{i}vel Superior (CAPES, Brazil) - Finance Code 001, and by the São Paulo Research Foundation (FAPESP), Brasil - Process Numbers 2022/07298-4, 2022/08335-0, 2023/07013-2.   
    MR acknowledges partial support from the Conselho Nacional de Desenvolvimento Científico e Tecnológico (CNPq, Brazil), Grant 315991/2023-2.
	RPM acknowledges support from the Villum Investigator program supported by the VILLUM Foundation (grant no. VIL37766) and the DNRF Chair program (grant no. DNRF162) by the Danish National Research Foundation and the European Union’s Horizon 2020 research and innovation programme under the Marie Sklodowska-Curie grant agreement No 101131233. 
	
	%%%%%%%%%%%%%%%%%%%%%%%%%%%%%%%%%%%%%%%%%%%%%%%%%%%%%%%%%%%%%%%%%%%%%%%%%%%%%%%%%%%%%%%%%%%%%%%%

	%\bibliographystyle{apsrev4-2}
	\bibliography{rz-blackhole}
	
\end{document}